\documentclass[sigconf, nonacm]{acmart}
\newcommand\vldbdoi{XX.XX/XXX.XX}

\newcommand\vldbvolume{XX}
\newcommand\vldbissue{X}

\newcommand\vldbavailabilityurl{}
\newcommand\vldbpagestyle{plain} 

\usepackage{balance}

\usepackage{algorithm}
\usepackage{algorithmicx}
\usepackage[noend]{algpseudocode}
\usepackage{graphicx}
\usepackage{textcomp}
\usepackage{xspace}
\usepackage{xifthen}
\usepackage{xargs}
\usepackage{hyperref}
\usepackage{regexpatch}
\usepackage{todonotes}
\usepackage{multirow}
\usepackage{mathtools}
\usepackage{comment}
\usepackage{etoolbox}
\usepackage{subcaption}
\usepackage{enumitem}
\usepackage{listings}             
\usepackage{textcomp}
\usepackage{booktabs}
\usepackage[normalem]{ulem}

\let\oldsection\section
\let\oldsubsection\subsection

\renewcommand{\section}{\vspace{-0.2cm}\oldsection}
\renewcommand{\subsection}{\vspace{-0.2cm}\oldsubsection}

\lstdefinestyle{customc}{
    language=C,
    keywordstyle=\bfseries\color{green!40!black},
    commentstyle=\itshape\color{purple!40!black},
    identifierstyle=\color{blue},
    stringstyle=\color{orange},
	basicstyle=\scriptsize\ttfamily,
	numbers=left,
	numberstyle=\tiny,
	numbersep=5pt,
	frame=lines,
	breaklines=true,
	prebreak=\raisebox{0ex}[0ex][0ex]{\ensuremath{\hookleftarrow}},
	showstringspaces=false,
	upquote=true,
	tabsize=2,
}

\lstdefinestyle{customsql}{
    language=SQL,
    keywordstyle=\bfseries\color{green!40!black},
    otherkeywords={STD},
    commentstyle=\itshape\color{purple!40!black},
    identifierstyle=\color{blue},
    stringstyle=\color{orange},
	basicstyle=\footnotesize\ttfamily,
	numbers=left,
	numberstyle=\tiny,
	numbersep=5pt,
	frame=lines,
	breaklines=true,
	prebreak=\raisebox{0ex}[0ex][0ex]{\ensuremath{\hookleftarrow}},
	showstringspaces=false,
	upquote=true,
	tabsize=2,
}

\makeatletter
\xpatchcmd{\@todo}{\setkeys{todonotes}{#1}}{\setkeys{todonotes}{inline,#1}}{}{}
\makeatother

\newcommand*\circled[1]{\tikz[baseline=(char.base)]{
    \node[shape=circle,draw,inner sep=1pt] (char) {#1};}}

\newcommand{\sname}{Relational Memory\xspace}

\newcommand\sysNameLong[0]{Relational Memory Engine\xspace}
\newcommand\sysName[0]{RME\xspace}

\newcommand{\Plimvar}{Ephemeral\xspace}
\newcommand{\plimvar}{ephemeral\xspace}

\newcommand\Paragraph[1]{\vspace{0.02in}  \noindent \textbf{#1.}}
\newcommand\Paragraphit[1]{\vspace{0.02in}  \noindent \textit{#1.}}
\newcommand\new[1]{{#1}}

\begin{document}

\title[{\emph \sname}: Native In-Memory Accesses on Rows and Columns]{{\emph \sname}: Native In-Memory Accesses on \\Rows and Columns}

\author{Shahin Roozkhosh}
\affiliation{\institution{\small Boston University\country{USA}}}
\email{shahin@bu.edu}

\author{Denis Hoornaert}
\affiliation{\institution{\small TU M\"unchen\country{ Germany}}}
\email{denis.hoornaert@tum.de}

\author{Ju Hyoung Mun}
\affiliation{\institution{\small Boston University\country{USA}}}
\email{jmun@bu.edu}

\author{Tarikul Islam Papon}
\affiliation{\institution{\small Boston University\country{USA}}}
\email{papon@bu.edu}

\author{Ahmed Sanaullah}
\affiliation{\institution{\small Red Hat\country{USA}}}
\email{asanaull@redhat.com}

\author{Ulrich Drepper}
\affiliation{\institution{\small Red Hat\country{USA}}}
\email{drepper@redhat.com}

\author{Renato Mancuso}
\affiliation{\institution{\small Boston University\country{USA}}}
\email{rmancuso@bu.edu}

\author{Manos Athanassoulis}
\affiliation{\institution{\small Boston University\country{USA}}}
\email{mathan@bu.edu}

\begin{abstract}

Analytical database systems are typically designed to use a column-first data layout to access only the desired fields. On the other hand, storing data row-first works great for accessing, inserting, or updating entire rows. Transforming rows to columns at runtime is expensive, hence, many analytical systems ingest data in row-first form and transform it in the background to columns to facilitate future analytical queries.
\emph{How will this design change if we can always efficiently access only the desired set of columns?}

To address this question, we present a radically new approach to data transformation from rows to columns. We build upon recent advancements in embedded platforms with re-programmable logic to design \emph{native in-memory
access on rows and columns}. 

Our approach, termed \emph{Relational Memory}, relies on an FPGA-based 
accelerator that sits between the CPU and main memory and transparently 
transforms base data to any group of columns with minimal overhead at 
runtime. This design allows accessing any group of columns as if it 
already exists in memory. 
We implement and deploy \emph{Relational Memory} in real hardware, and we 
show that we can access the desired columns up to 1.63$\times$
faster than accessing them from their row-wise counterpart, while 
matching the performance of a pure columnar access for low projectivity,
and outperforming it by up to 1.87$\times$ as projectivity (and tuple reconstruction cost) increases. 
Moreover, our approach can be easily extended to support 
offloading of a number of operations to hardware, e.g., selection, group 
by, aggregation, and joins, having the potential to
vastly simplify the software logic and accelerate the query execution.

\end{abstract}

\maketitle

\pagestyle{\vldbpagestyle}
\begingroup
\renewcommand\thefootnote{}\footnote{\noindent
This work is licensed under the Creative Commons BY-NC-ND 4.0 International License. Visit \url{https://creativecommons.org/licenses/by-nc-nd/4.0/} to view a copy of this license. For any use beyond those covered by this license, obtain permission by emailing \href{mailto:info@vldb.org}{info@vldb.org}. Copyright is held by the owner/author(s). Publication rights licensed to the VLDB Endowment. \\
\raggedright Proceedings of the VLDB Endowment, Vol. \vldbvolume, No. \vldbissue\ %
ISSN 2150-8097. \\
\href{https://doi.org/\vldbdoi}{doi:\vldbdoi} \\
}\addtocounter{footnote}{-1}\endgroup

\ifdefempty{\vldbavailabilityurl}{}{
\vspace{.3cm}
\begingroup\small\noindent\raggedright\textbf{PVLDB Artifact Availability:}\\
\url{\vldbavailabilityurl}.
\endgroup
}

\vspace{-0.2in}
\section{Introduction}
\label{sec:intro}

\Paragraph{OLTP vs. OLAP vs. HTAP}
Over the past few years, large-scale real-time data analytics has soared in popularity as more and more applications need to analyze fresh data. This has been exacerbated by new technological trends like 5G, Internet-of-Things, and the advent of cloud computing as an always-on data platform \cite{Cisco2018,Gartner2017}. This leads to the need for systems that can perform both Online Transactional Processing (OLTP) and Online Analytical Processing (OLAP), known as Hybrid Transactional/Analytical Processing (HTAP)~\cite{Ozcan2017}.
However, OLTP and OLAP systems adopt very different designs. OLTP systems are generally optimized for write-intensive workloads aiming to support high-volume point queries using indexes. In contrast, OLAP systems are optimized for read-only queries that access large amounts of data. Recent efforts for HTAP systems have been bridging OLAP and OLTP requirements by maintaining multiple copies of data in different
formats~\cite{Barber2015,Ramamurthy2003} or converting data between different layouts~\cite{Appuswamy2017,Arulraj2016,Lahiri2013,May2017,Shamgunov2014}.

\begin{figure}[t]
	\centering
	\includegraphics[width=0.42\textwidth]{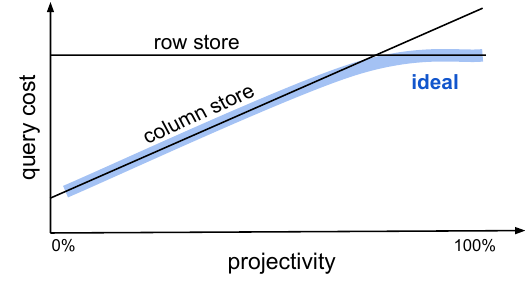}
\vspace{-0.2in}
	\caption{Row-wise accesses have constant cost, while columnar accesses have higher cost for higher projectivity. Ideally, the cost should be the minimum of the two. }
	\label{fig:intro-ideal}
\vspace{-0.2in}
\end{figure}

\Paragraph{Data Layout}
Efficient access to useful data is a key design goal for all data-intensive systems. Typically, this manifests as finding the optimal data layout, with the traditional schism being \emph{row-stores} vs. \emph{column-stores}.  {\bf Transactional systems} employ row-stores, meaning that the physical organization of data items in memory is structured in sequential rows. Row-stores are ideal for queries that either update the contents of a single row, append a new row, or focus on the full information contained in several rows. On the other hand, most {\bf analytical systems} store data in a columnar fashion that supports fast scan of specific attributes. Column-stores focus on grouping together the same attribute of different rows, hence allowing for efficient analytical querying, i.e.,  where typically information from multiple rows is aggregated~\cite{Abadi2013}.

In order to bridge the analytical and transactional requirements, many HTAP systems use a single architecture that ingests data in row-format, and once data hit the disk, converts them into a columnar format~\cite{Ozcan2017}. By doing so, HTAP systems fuse the pipelines of ingesting data and performing analytics, leading to a single data-store maintaining the fresh data and efficient analytics on that same data set.
Analysis of systems with adaptive layouts, like H$_2$O~\cite{Alagiannis2014}, Hyper~\cite{Kemper2011}, Peloton~\cite{Arulraj2016}, and OctopusDB~\cite{Dittrich2011} showed that every query has an optimal layout which is neither a column-store nor a row-store. However, defining and maintaining multiple layouts carries a large amount of extra complexity, which leads to runtime inefficiency arising from heavy book-keeping overheads. The additional complexity of the codebase also results in scalability and maintainability issues.

\begin{center}
  \textbf{\emph{
    What if the optimal layout was always available?
  }}
\end{center}

In other words, ``\emph{what if the underlying hardware allows us to access only the desired groups of columns while the data is stored in memory as a row-store?}'' Typically, when using a row-store, we always have to fetch the entire row through the memory hierarchy, irrespectively of the projectivity of the query. 
On the other hand, column-stores allow us to bring only the desired columns 
with increasing tuple reconstruction cost as we increase projectivity. This leads 
to higher query latency when projectivity is close to 100\%~\cite{Alagiannis2014}. 
The resulting expectation of query cost as a function of projectivity is shown in Figure~\ref{fig:intro-ideal}. Ideally, we would want to pay only for the useful data and have negligible tuple reconstruction cost. This could be achieved by seamlessly switching between column-store and row-store depending on what constitutes the better choice, given the projectivity of the query at hand. 
Prior work on systems that support adaptive layouts via code generation~\cite{Alagiannis2014,Karpathiotakis2014,Kemper2011} fix a base storage that uses either a row-store or a column-store as a starting point, and the adaptive layout is generated via copying only the relevant data. In turn, this approach creates the need for managing and invalidating these copies upon updates.

\begin{center}
  {\emph{In this paper, we propose a paradigm shift. We propose a
      novel hardware design for a data reorganization engine that (1)
      can be implemented in existing commercial platforms, (2) is
      capable of intercepting CPU-originated memory requests, and (3)
      of producing responses where the supplied data items are always
      transparently arranged in the most efficient layout, whilst (4)
      the source data tables are always stored in physical memory
      according to the same format---i.e., as a row-store. We show how
      to integrate this new hardware design with clean abstractions
      for data system developers to benefit from it without having to
      redesign the entire database engine.}}
\end{center}

\Paragraph{\sname}
To offer native access to any group (whether it be continuous or non-contiguous) of columns without overhead, we create 
new specialized hardware that acts as an on-the-fly data transformer from 
rows stored in memory to any group of columns shipped through the memory 
and cache hierarchy toward the processor. We utilize commercially available 
systems-on-chip (SoCs) that include programmable logic (PL) and a traditional 
multi-core processing subsystem (PS) on the same chip. These PS-PL SoCs allow 
the design and deployment of resource management primitives and create usable 
proof-of-concept prototypes to assess performance benefits with realistic 
applications. 

Specifically, we capitalize on recent advancements in reprogrammable
hardware~\cite{Roozkhosh2020} that allow us to implement programmable logic 
\emph{between the memory and the processor}. To ensure ease of programmability, 
we do not directly expose the specialized hardware to the data system engineer. 
Instead, we expose a simple abstraction that allows them to request the desired 
column groups and transparently use the underlying machinery. We refer to the 
ability to provide an on-the-fly representation of the data that optimizes 
relational operators as \emph{\sname}.

\Paragraphit{Ephemeral Variables} In a database management system (DBMS) implementation, every relational table loaded in memory is accessible through a variable. By default, this points to the base row-oriented representation of the data, tailored for accessing entire rows and updating or inserting data. Different analytical queries, however, might require a different subset of the relation's columns. To support this, we introduce \emph{ephemeral variables}, a particular type of variable that identifies a specific subset of columns to access. These variables are never instantiated in the main memory. Rather, upon accessing such a variable, the underlying machinery is set in motion and generates on-the-fly a projection of the requested columns according to the format that maximizes data locality. 

The philosophy behind \sname pivots on three main points:
(1) pushing relational operators closer to data storage; (2) reorganizing
and compacting data items before they are moved toward CPUs to improve locality; and (3) relying on traditional CPUs for data processing once good locality has been achieved. Operating closer to the data also introduces opportunities to exploit the inherent parallelism of memory cells---e.g., by issuing outstanding parallel requests to separate DRAM banks.
Note that hardware prefetching can benefit from the memory cells parallelism as long as the accesses follow a sequential logic. \sname, however, has semantic knowledge that helps to perform operations out of sequence and still exploits the inherent memory parallelism. 
Reorganizing data to improve locality minimizes the waste of constrained CPU cache estate. In turn, this translates to better efficiency for the query at hand and lower cache pollution. Lastly, we enable seamless integration with existing data management systems by limiting our design to data reorganization while relying on CPUs to implement arbitrarily complex analytics.


\begin{figure}[t]
	\centering
	\includegraphics[width=0.48\textwidth]{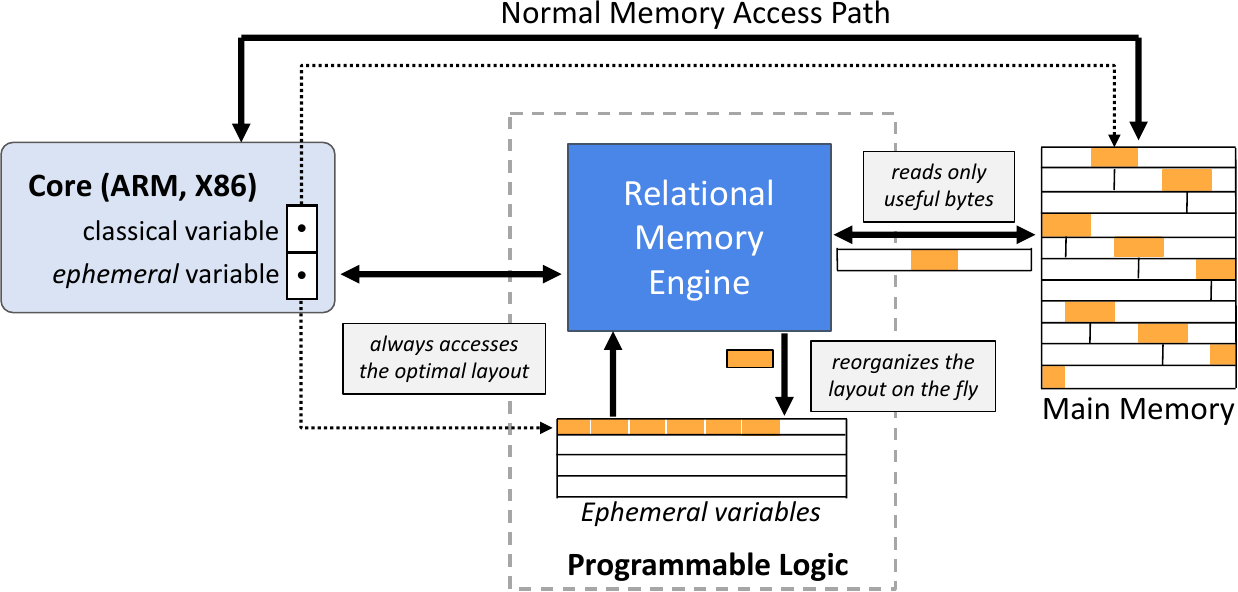}
	\vspace{-0.25in}
	\caption{Architecture of the proposed \sysNameLong. The engine
          pushes relational operators closer to data and provides the
          optimal layout for a given workload via on-the-fly data
          reorganization.}
	\label{fig:intro-relmem}
	\vspace{-0.2in}
\end{figure}

Figure~\ref{fig:intro-relmem} shows a high-level diagram of the proposed design.  The \sysNameLong (\sysName) is located in the programmable logic between the memory and the processor. Upon receiving a request that triggers it, the engine transforms data rows to any desired combination of columns on-the-fly. The processor directly accesses data in the optimal layout through pointers to \emph{ephemeral variables} that triggered the transformation.

\Paragraph{Contributions} 
To the best of our knowledge, \sname is the first hardware/software co-design that allows native access to both rows and column-groups over data stored in a row-wise format in memory. Our prototype FPGA implementation supports projection and lays the groundwork for pushing more functionality, i.e., selection, aggregation, group by, and join pre-processing. Pushing projection to the hardware creates opportunities for fundamental changes in software/hardware co-design for data systems by \emph{enabling systems to have native access to any data layout}.
In turn, this leads to better cache utilization and eventually paves the way towards a unified HTAP architecture even in the presence of queries with very different access patterns and requirements, essentially
taking a fundamental step towards truly HTAP systems. Our work offers the following concrete contributions. 

\vspace{-0.02in}
\begin{itemize}[leftmargin=1.3em, itemsep=0.25em]
  \item We present \emph{\sname}, a novel SW/HW co-design paradigm for general-purpose query engines;
  \item \sname ensures that every query has always access to the optimal data layout;
  \item We propose \emph{ephemeral variables}, a simple and lightweight abstraction to use \sname;
  \item We implement an FPGA proof-of-concept prototype that already demonstrates the viability and the potential impact of our design; 
  \item We experimentally show that the implemented \sysName performs native accesses to groups of columns, as if the ideal layout is available in memory with no extra cost to transform rows to columns, leading to higher cache efficiency. 
\end{itemize}

\section{Background}
\label{sec:background}

We now introduce the key concepts necessary to explain the design of \sname.
First, we briefly introduce the nuts and bolts of the FPGA technology and the
typical organization of PS-PL 
platforms. Next, we discuss the
Programmable Logic In-the-Middle (PLIM) approach that this work builds
upon. Lastly, we discuss the key principles for data organization in
database systems.

%


\vspace{-0.01in}
\subsection{FPGA Background}
\vspace{-0.03in}
\Paragraph{Field-Programmable Gate Arrays} FPGAs are programmable devices
that can be configured to synthesize hardware functional blocks~\cite{kuon2008fpga, trimberger2012field}. FPGAs
are becoming increasingly popular in modern platforms because of their
high parallelism, reconfigurability, specializability, and power
efficiency. In comparison to more traditional CPUs, co-processors, and
GPUs, they do not rely on the execution of a set of instructions.
Instead, using a \emph{bitstream} mapping of a synthesized version of the
logic to its internal components, they are capable of directly
emulating the logic of any digital circuit. This unique capability
makes them attractive in many applications as such tailored designs
provide high performance because of specific data manipulation, near
perfect locality of the data, and many levels of parallelism (e.g.,
pipelining). For this reason, FPGA technology has been widely used to
implement specialized accelerators.


An FPGA device is organized as a large number of programmable logic
blocks surrounded by an interconnect fabric. \textit{Lookup tables}
(LUTs) are the main building block in programmable logic. Each LUT is
essentially an $n$-input, 1-output table to be configured with an
arbitrary boolean function~\cite{rose1993architecture, ahmed2004effect}. Nowadays, multi-output LUTs are also
available~\cite{mlut}. Multiple LUTs can be connected using the
configurable interconnect fabric. In addition to the logic circuits,
an FPGA has both a small memory in the forms of registers or
flip-flops and a larger local memory implemented as Block RAM, or BRAMs
for short, and can access large but slow \emph{off-chip} memories
through the DRAM controllers. Internal memory can reach TB/s scale
bandwidth with sub-microsecond latency, whereas off-chip memories'
bandwidth can reach GB/s~\cite{DBLP:conf/fpl/MinHEXCH19}.

\begin{figure}[]
  \centering
  \includegraphics[width=0.4\textwidth]{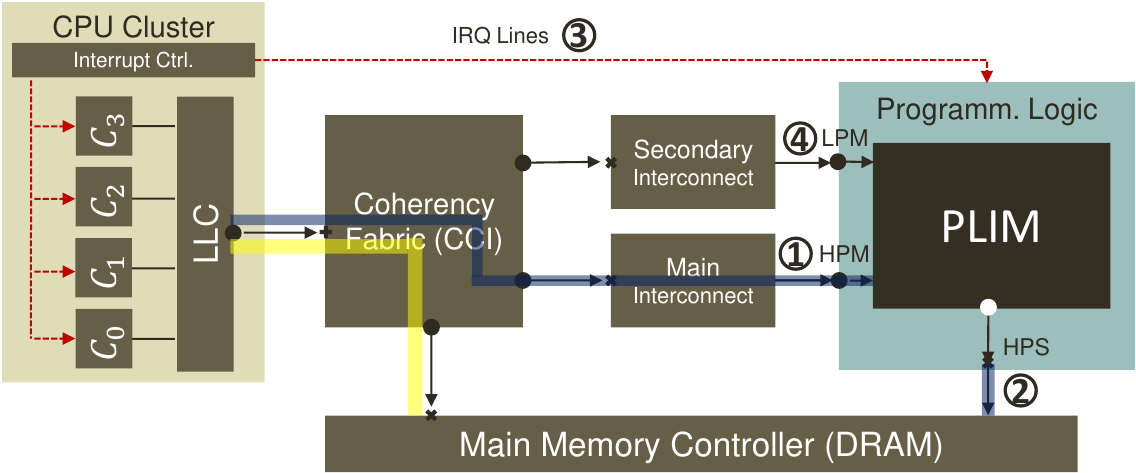}
\vspace{-0.1in}
  \caption{PLIM module instantiation on a PS-PL platform.}
  \label{fig:fpga}
\vspace{-0.15in}
\end{figure}

\Paragraph{PS-PL Platforms}
Recent years have seen the advent of PS-PL platforms, heterogeneous
System-on-Chip (SoC) where a traditional system (referred to as
\emph{PS-side}) is associated with a tightly integrated piece of
programmable-logic, i.e., an FPGA (referred to as PL-side).  The
interest surrounding these platforms has gained some momentum with the
proliferation of available models produced by
Intel~\cite{intel_stratix10}, Xilinx~\cite{zynq_ultrascale}, ETHZ~\cite{enzian2020cidr,ETHZ2021}, and Microsemi with PolarFire
SoC~\cite{microsemi_polarfire}. As shown in Figure~\ref{fig:fpga}, the
PL domain can communicate through high-performance PS-PL communication
interfaces(\circled{1}, \circled{2}, \circled{4}), or Interrupt lines
(\circled{3}) with the rest of the system. On-chip communications
are carried out using a high-performance, synchronous, high-frequency,
multi-primary/secondary communication interface between
functional blocks.

These platforms use the popular, open specification, and
widely adopted \emph{Advanced eXtensible Interface} (AXI)
protocol~\cite{ARM-AXI}. The latter supports asynchronous
\textbf{read} and \textbf{write} transactions through dedicated
channels operating in parallel between a primary
(a processor) and secondary (a memory device). In addition,
the protocol allows the primaries to emit multiple
outstanding transactions. Each sequence of transactions is
identifiable through a given ID~\cite{ARM-AXI}. 

\subsection{Programmable Logic in The Middle}

Traditionally, in PS-PL platform, the PL-side is used to map
hardware accelerators that work in a load-unload fashion.  However,
Roozkhosh et al.~\cite{Roozkhosh2020} have prompted a shift
in paradigm with the introduction of the Programmable Logic in the
Middle (PLIM) approach.  This approach differentiates itself from the
others in light of the way it considers and uses the PL-side.  In
fact, this approach advocates the use of the PL-side as a secondary
route to main memory that can entirely or partially replace the
\emph{normal} route.  As illustrated in Figure~\ref{fig:fpga}, instead
of using the normal data path (highlighted in yellow), CPU
traffic can be redirected through the PL-side before reaching the main
memory (highlighted in blue).

The principal advantage of PLIM resides in its ability to intercept,
inspect and manipulate any PS-originated memory transactions before
they reach the main memory.  This capability of inspecting memory
traffic at the granularity of individual transactions has already been
leveraged to address and tackle bedeviling problems.  For instance,
PLIM demonstrates how by simply manipulating each transaction address,
memory fragmentation introduced by address coloring can be mitigated~\cite{Roozkhosh2020}.
The same authors have also shown that the same type of module can be
integrated into a wider framework in order to address the problem of
memory traffic scheduling~\cite{Hoornaert2021}.  Interestingly, PLIM
modules have also been used to highlight the possibility of
memory-based on-chip denial-of-service attacks from remote cores under
special conditions~\cite{Hoornaert2021b}.

\vspace{-0.01in}
\subsection{Data Layouts}
\vspace{-0.02in}
A key decision for any data system is the employed \emph{data layout},
which is tightly connected with the type of workloads it primarily
targets.  In general, there are two extremes according to which DBMS's
store data: the $n$-ary storage model (\emph{row-stores}) and the
decomposition storage model (\emph{column-stores}).  Row-stores
follow the volcano-style processing model where data is organized as
tuples, and all the attributes for each tuple are stored
sequentially~\cite{Padmanabhan2001, Hellerstein2007}.  This design
allows superior performance for OLTP workloads~\cite{Ailamaki1999}
since the queries in transactions generally tend to operate on
individual tuples.
Traditional DBMS like Oracle~\cite{Alexander2005}, IBM
DB2~\cite{Cain2011}, SQL Server~\cite{Agrawal2004} follow this
paradigm.
In contrast, column-stores are organized as columns following the
decomposition storage model.  They process data one column at a time,
hence are better suited for OLAP workloads~\cite{Abadi2013} where the
queries tend to operate on multiple tuples but only access a small
fraction of the attributes.
Most contemporary data systems like Vertica~\cite{Lamb2012}, Actian
Vector (formerly Vectorwise~\cite{Zukowski2012}),
MonetDB~\cite{Boncz2005}, Snowflake~\cite{Dageville2016} use columnar
storage.  Even traditional row-stores have developed new variants
that allow to store data in columnar format~\cite{Lahiri2015,
  Barber2015, Larson2013}.  Because of these design principles, OLTP
queries are costly in column-stores while OLAP queries are expensive
for row-stores.  Systems targeting hybrid workloads support
\emph{hybrid layouts} in the form of column-groups via the flexible
storage model~\cite{Arulraj2016} or adaptive
layouts~\cite{Alagiannis2014,Dittrich2011,Kemper2011}.  These systems
dynamically adapt the storage layout depending on the workload by
keeping the same data in different layouts and by converting the data
between row and columnar formats for transactions and analytics,
respectively.  Because of these conversions and multiple layouts,
these systems generally have high complexity, high materialization
cost, and heavy book-keeping overheads.

\section{\sname}
\label{sec:relmem}


We now present the high-level design of our \sysNameLong (\sysName) and 
the interface that allows its transparent use.

\Paragraph{Implementing Relational Algebra Operators in Hardware}
The core innovation introduced with the \sname paradigm is the
implementation of relational algebra operators in hardware to achieve
transparent near-data processing. \sname represents a novel
software-hardware co-design approach with a clean architecture and a
simple abstraction that is exposed to the application level.

The fundamental physical relational operators are projection,
selection, sorting, aggregation, group-by, and join. In this paper, we
focus on projections that require the capability to fetch a subset of
the data residing in memory (projecting the desired
columns). Realizing near-data projection lays the groundwork for
pushing the highest degree of processing to the hardware as long as
data movement represents the performance bottleneck.

Current state-of-the-art systems that support hybrid layouts create the 
desired column-groups in software, therefore, the 
data has to pass through the memory hierarchy and get copied in order to 
create the desired layout. On the contrary, 
we propose to make \emph{any layout available on the fly} by creating a special memory address
that is an alias of the original data pointer---a.k.a. an \textbf{ephemeral variable}.
CPU accesses to ephemeral variables are intercepted by our \sysName that
constructs a response to each memory request. The response payload is produced
by packing only useful data. Therefore, from a CPU's standpoint, the data appears as
always structured according to the optimal layout for the query at hand. Two key 
benefits are that (1) we do not duplicate data in memory as the ephemeral variables provide
a reorganized view of the original data; and (2) unlike traditional hardware
accelerators, the CPU can immediately access partial results without having to wait for
the \sysName to complete a full pass over the original data. In fact, only memory
requests for non-ready data chunks are stalled by the \sysName.

\begin{figure}
    \centering
    \vspace{-0.3cm}
    \includegraphics[width=0.95\linewidth]{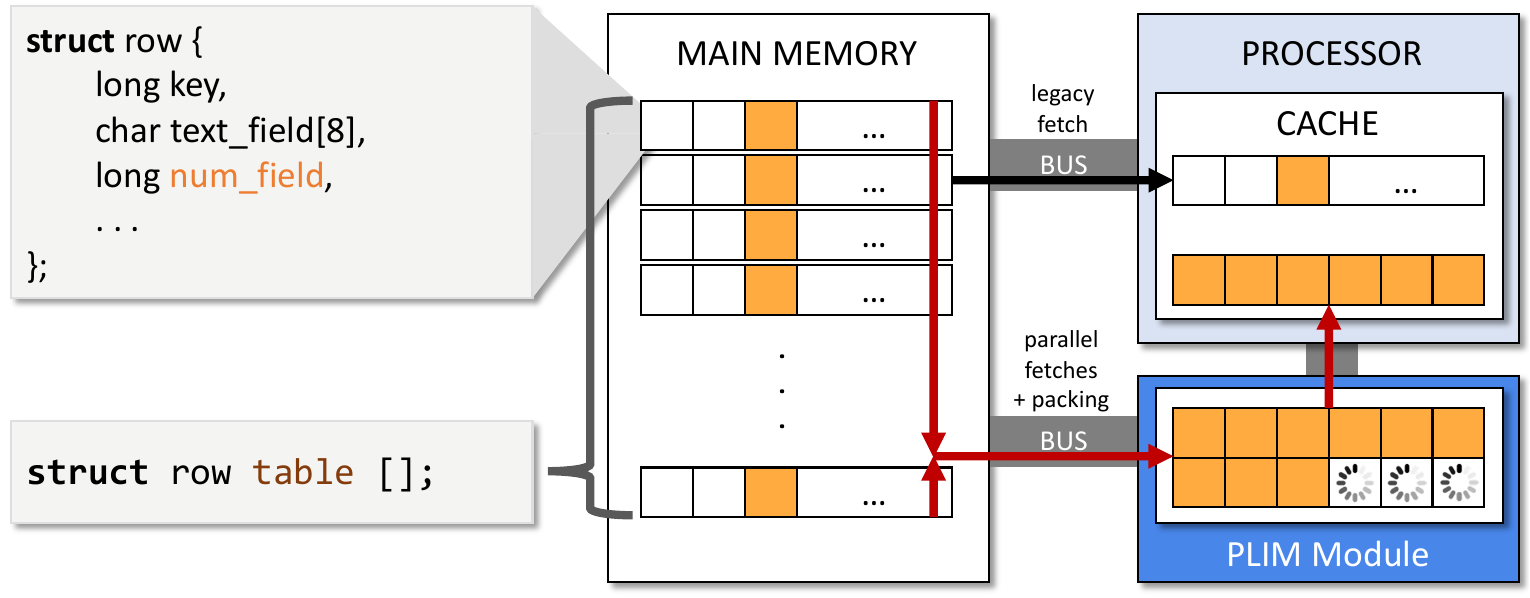}
    \vspace{-0.1in}
    \caption{Example of in-line data transformation using the proposed 
    \sname{}.}
    \label{fig:plim_reorder}
    \vspace{-0.2in}
\end{figure}

\Paragraph{A Low-Level Example}
One can think of a row in a 
database table as a struct of the type \mbox{\texttt{struct row table}} 
as shown in Figure \ref{fig:plim_reorder}.
A row-oriented table is simply an array of rows of the type 
\mbox{\texttt{struct row table []}}. If one wants to access only 
the numeric field (\texttt{num\_field}) from all the rows, i.e., 
perform a column access, this creates a stride-pattern 
data access, where 8 bytes are accessed every few hundred bytes 
of data. This is inefficient because (1) each new row always pulls 
an entire cache line from memory; (2) the large strides are not 
handled well by hardware prefetchers; and (3) in general, more 
data is transported from main memory to the processors than what 
is strictly required for the requested type of access. In the 
considered example, with a cache line size of 64~bytes, only 1/8 
of a cache line is utilized.

A column-store optimizes these types of accesses by 
storing each attribute separately, hence allowing for accessing only 
the numerical field through an array of the type: 
\texttt{long num\_field\_array []}. Indeed, in this case, only data 
items strictly required for the final computation are transported 
from main memory, resulting in a highly localized access pattern. However, 
this comes at the cost of having an inefficient layout for insertion and
deletion, and paying increasing tuple reconstruction cost with higher
projectivity.

\Paragraph{Near-Data Projection}
To offer contiguous access to a specific column (or group of columns), 
we design our \sysName leveraging the PLIM paradigm. The PLIM
\cite{Roozkhosh2020} design is conceptually similar to
Processing-In-Memory (PIM)~\cite{Loh2013} and Near-Memory
Processing~(NMP) \cite{Balasubramonian2014} as they all execute logic
close to memory.  The key innovation of the proposed \sysName is that
it creates data that does not exist in main memory, which the CPU can
use transparently as if it exists in main memory. \sysName can be
implemented in FPGA in a PS-PL platform. As we demonstrate in
Section~\ref{sec:eval}, our FPGA prototype already provides a significant
performance advantage. Nonetheless, we envision that additional
benefits can be unlocked by embedding a similar design \emph{within}
the memory controller itself.
%
%
The \sysName creates memory \emph{aliases} to expose non-contiguous
content as if it were contiguous.  In other words, the module enables
accessing the same content in main memory under different strides, but
always as if it were stored contiguously from the perspective of the
main CPUs. This is drastically different compared to traditional
scatter-gather strategies~\cite{Seshadri2015} initiated by typical DMA-capable
accelerators (e.g., SIMD processors) because data
transformation is performed in-line with the instruction stream, with
fine-grained information on the exact byte-wise location of data items that are useful
for the computation at hand; and because it allows predicting and exploiting data reuse
across processing micro- and macro-phases. The \sysName receives as
input the intended access stride of the query (that maps to the
physical addresses of the columns to be accessed). It then issues parallel
main memory requests for the target data. Finally, it assembles
multiple entries in a single \emph{packed} cache line to be sent to the processor.

This abstraction creates \emph{non-materialized in-memory aliases of 
column-groups} that push up to the memory hierarchy arbitrary subsets 
of columns in \emph{dense memory addresses} from the cache memory perspective.
That way, both efficient column-oriented and row-oriented accesses can be supported
while minimizing CPU cache pollution with unnecessary attributes.

\Paragraph{\Plimvar Variables} 
In order to provide control to 
use and initialize the proposed hardware, we propose a 
lightweight software/hardware interface. Specifically, 
to use \sname, the data system developer
shall use a new type of memory pointers termed
\emph{\plimvar variables}, which do not correspond to a 
real main memory location. Any CPU access on
\plimvar variables (that leads to a cache miss) is routed 
to and satisfied by the PL. To understand the semantics,
the expected content, and the versatility of \plimvar 
variables, let us consider a concrete example.

Suppose that a full relational table is loaded in memory and structured 
as a classic 2-D array, as previously discussed. 
For instance, the table under analysis corresponds to 
the array \texttt{struct row table[]}, where each row is defined as:

\vspace{-0.05in}
\lstset{escapechar=@,style=customc}
\begin{lstlisting}[caption={C-style relational table row definition.}, 
label=lst:dbrow]
struct row {
    long key;               /* 8 bytes*/
    char text_fld1 [8];     /* 8 bytes */
    char text_fld2 [12];    /* 12 bytes */
    char text_fld3 [20];    /* 20 bytes */
    char text_fld4 [16];    /* 16 bytes */
    long num_fld1;          /* 8 bytes */
    long num_fld2;          /* 8 bytes */
    long num_fld3;          /* 8 bytes */
    long num_fld4;          /* 8 bytes */
    long num_fld5;          /* 8 bytes */
};
\end{lstlisting}

\noindent In order to have direct access to a single column, or to a group of 
columns, we create an \plimvar variable that is registered 
with the \sysName. Accessing the newly created \plimvar 
variable from the CPU's perspective is equivalent to having direct 
access to a subset of columns with a packed view of the relevant fields. 
Following our example, to access only columns 
\texttt{num\_fld1}, \texttt{num\_fld3}, and \texttt{num\_fld4},
we create an \plimvar variable of the type:

\vspace{-0.05 in}
\lstset{escapechar=@,style=customc}
\begin{lstlisting}[caption={C-style \plimvar type definition.}, 
label=lst:dbrow_proj]
struct column_group_1 {
    long num_fld1;         /* 8 bytes */
    long num_fld3;         /* 8 bytes */
    long num_fld4;         /* 8 bytes */
};
\end{lstlisting}

\noindent The \plimvar variable with type \texttt{column\_group\_1}
provides access to the three desired columns as a contiguous 
array. This comes with three advantages. First, only useful 
information is propagated through the cache hierarchy, dramatically 
reducing cache pollution and working-set size, and thus improving 
cache reuse and locality. Second, \sysName orchestrates 
data accesses to main memory in a way that is DRAM 
structure-aware to maximize throughput---much like a DMA would. 
Third, having turned a stride access, potentially spanning  
multiple pages, into a sequential pattern over a smaller buffer 
greatly improves the effectiveness of CPU-side prefetching.

\Paragraph{The Lifetime of a Memory Access} 
Here, we demonstrate the lifetime of a memory access targeting an
\plimvar variable, through a sample analytic query:

\vspace{-0.05in}
\lstset{escapechar=@,style=customsql}
\begin{lstlisting}[caption={Sample projection+aggregation query.}, label=lst:query-sql]
SELECT sum(num_fld1 * num_fld4) 
    FROM the_table
    WHERE num_fld3 > 10; 
\end{lstlisting}

\noindent This query only requires accessing three out of the ten columns,
and can be evaluated using an \plimvar variable as
follows:

\vspace{-0.05in}
\lstset{escapechar=@,style=customc}
\begin{lstlisting}[caption={Query logic in C language.}, label=lst:query-c]
struct row the_table[];
/* Autogenerated Code Block - START*/
struct column_group_1 {
    long num_fld1;
    long num_fld3;
    long num_fld4;
};
struct column_group_1 cg[] = @\label{line:grp_str_start}@
    register_var(the_table, num_fld1, @\label{line:reg}@
                num_fld3, num_fld4); @\label{line:grp_str_end}@
/* Autogenerated Code Block - END */
int sum = 0;
for (int i = 0; i < cg.length; i++) {
    if (cg[i].num_fld3 > 10) {
        sum += cg[i].num_fld1 * cg[i].num_fld4;
    }
}
\end{lstlisting}

\noindent Note that this is a simplified code snippet. To optimize for
performance, one can implement state-of-the-art approaches 
including \emph{predication}~\cite{Allen1983,August1997} to avoid branch 
misprediction and \emph{vectorization}~\cite{Boncz2008} to increase 
locality and computation efficiency. 
Orthogonally to those optimization strategies, we focus on minimizing 
data movement. \Plimvar variables fetch 
only relevant columns from memory leading to 
optimal cache utilization.
The moment an \plimvar variable is registered (Listing~\ref{lst:query-c}, line~\ref{line:reg}), the 
geometry of the access is defined (that is, the pattern of 
scattered accesses on the original data), and when the data 
is first accessed---i.e., when the statement \texttt{cg[0].num\_fld3 > 10} 
is evaluated---the \sysName starts projecting only the 
relevant columns. We present the design of \sysName in detail in Section~\ref{sec:design}.

\vspace{0.02in}
\section{Discussion}
\label{sec:discussion}
\vspace{-0.03in}

\Paragraph{Updates \& MVCC Transactions}
\new{
While \sname offers native access to both rows and columns, the base data are stored 
in-memory in a row-oriented format. We treat all ephemeral variables as 
\emph{read-only} columns or group-of-columns that accelerate analytical queries. 
Updates are handled by accessing the \emph{read/write} row-oriented base data. Specifically, 
new rows are appended in the base data. In order to support in-place updates and 
deletion we use two timestamp fields for every row, storing potentially multiple 
versions. The first timestamp is set when the row is
inserted and marks the beginning of its validity, and the second is set when the row
is deleted or replaced by a newer version, marking the end of its validity. Every time
an ephemeral variable is accessed, it generates the (group of) column(s) that contain
the rows that are valid at the time of the query.
Using the timestamp scheme discussed above, \sysName also support MVCC transactions 
via snapshot isolation. 
}

\Paragraph{Compression}
\new{
\sname natively supports dictionary and delta (frame of reference) encoding that is 
frequently used in state-of-the-art column-store 
systems~\cite{Abadi2006,Abadi2013,Zukowski2006a}. Note that both can be used in 
row-oriented data and hence, they can benefit any groups of columns requested by
\plimvar variables.
}

\new{
Another compression scheme used in column-store systems is run-length encoding 
(RLE)~\cite{Abadi2006}. Contrary to dictionary and delta encoding, RLE has an expensive 
decoding step and relies on the data being sorted. RLE achieves typically higher
compression rates, but is less frequently applicable, hence, typically, is not 
preferred over dictionary and delta compression~\cite{Zukowski2006a}.
}

\Paragraph{Indexes \& Execution Strategies}
\new{
Base data indexes on the row-major data can still be very useful when updating the data
(using the MVCC approach outlined earlier) and when we have a very selective query. 
Note that \sname revolutionizes the software design of query engines by offering
at the same time \emph{native} access to both columns and rows. Hence, at runtime,
the query optimizer can decide to execute one query with indexes and another query
with columns, alternating between a row-at-a-time and column-at-a-time execution
strategy depending on what is the best fit for each query. In this paper, we provide
the hardware infrastructure for this radical paradigm shift and showcase the 
opportunity of accessing any data layout on a per-query basis. Implementing a 
full-fledged hybrid query engine is left for future work.
}

\Paragraph{Fractured Mirrors without the Mirrors}
\new{
Essentially, in this paper, we take a big step from the fractured mirrors 
approach~\cite{Ramamurthy2002,Ramamurthy2003} to offer access to both a row-store and
a column-store version of the data, along with everything in-between (arbitrary groups 
of columns), \emph{without maintaining multiple copies of the data}. 
}

\newcommand{\profc}{\texttt{profctl}\xspace}
\newcommand{\profv}{\texttt{profvm}\xspace}
\newcommand{\staller}{Trapper\xspace}
\newcommand{\requestor}{Requestor\xspace}
\newcommand{\fetch}{Fetch Unit\xspace}
\newcommand{\fetchs}{Fetch Units\xspace}
\newcommand{\bypass}{Monitor Bypass\xspace}
\newcommand{\extcol}{Column Extractor\xspace}
\newcommand{\reader}{Reader\xspace}
\newcommand{\writer}{Writer\xspace}
\newcommand{\config}{Configuration Port\xspace}
\newcommand{\relacc}{\sysName}
\newcommand{\relcache}{Reorganization Buffer\xspace}
\newcommand{\addr}{\texttt{A}}
\newcommand{\id}{\texttt{ID}}
\newcommand{\data}{\texttt{RD}}


\section{H/W Design and Implementation}
\label{sec:design}


The goal of our design is to enable in-memory data storage in
a single format (i.e., row-stores) while offering a \emph{reorganized
  view} of the same data with ideal locality. To do so, we interpose a
PL engine between CPUs and main memory. The data organization in main 
memory never changes but the semantics of memory accesses performed by 
the CPUs are re-defined on the fly. The proposed \emph{data 
reorganization engine} goes under the name of \sysName. The engine uses 
knowledge
of the target relation's geometry to make the retrieved data appear as
if they were initially stored in compacted projections.

\begin{figure}[t!]
  \centering
	\includegraphics[width=0.42\textwidth]{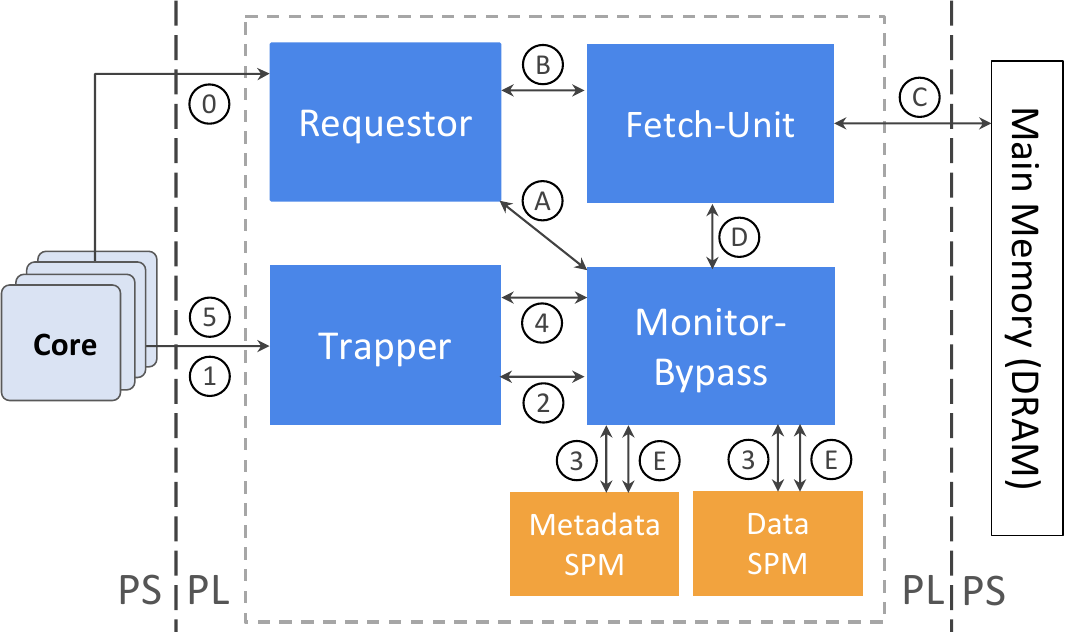}
  \vspace{-0.1in}
	\caption{Abstract overview of the RelBuffer components and interconnections with the PS-side.}
	\label{fig:design}
  \vspace{-0.2in}
\end{figure}

As depicted in Figure~\ref{fig:design}, our \sysName is comprised of
six modules: (1) \emph{Configuration Port}, (2) \emph{Monitor Bypass},
(3) Trapper, (4) \emph{Requestor}, (5) \emph{Fetch Unit}, and (6)
\emph{Relational Buffers}. The \sysName interacts with the PS through
two primary and one secondary AXI port. This section provides a
bird-eye's view of the \sysName operating mode and the role of each of
its sub-components.

\Paragraph{Configuration Port}
The architecture features a configuration port that enables the DBMS
to specify the location and geometry---that is, the database tuple width, tuple count along with size and positions of the requested columns---of the targeted database table at
runtime. This enables the \sysName to be runtime-configurable and
hence to be used for multiple queries. As listed in
Table~\ref{tab:configuration-port}, we distinguish seven of group of parameters:
(1) the \emph{row size} $R$ represents the size in bytes of the target table row; 
(2) the \emph{row count} $N$ is the total number of rows composing the target table; 
(3) the \emph{software reset} $SW$ enables the software layer to enforce a reset of the \sysName; 
(4) the \emph{enabled columns} $Q$ indicates the amount of columns of interrest (max 11); 
(5) the \emph{column width} $C_{A_j}$ captures the width in bytes of the $j$-th column of interest; 
(6) the \emph{row offset} $O_{A_j}$ represents the offset in bytes of the $j$-th column of interest from the offset of the previous column of interest (i.e., $j-1$). Therefore, the offset of column $j$ from the beginning of the current row can be computed as $\sum_{k=0}^{j} O_{A_k}$; and 
(7) the frame number $F$.
In our proof-of-concept implementation, the maximum column width
is set to 64~bytes---one full cache-line---and up to 11 non-contiguous columns can be specified.
These limitations are not fundamental to our design but only an implementation artifact.


\Paragraph{\bypass}
The \emph{\bypass} is the most central module in the proposed architecture. It is in charge of managing, synchronizing, and controlling the other modules. Specifically, it is responsible for (1) interacting with the \staller to acknowledge and answer incoming requests, (2) collecting data coming from the \fetchs, (3) monitoring the completion and availability status of each chunk of reorganized data, and (4) activating the \requestor when access to the first chunk of data is detected after the \sysName's reconfiguration.

\Paragraph{\staller}
The \emph{\staller} module is in charge of interfacing the
\sysName's internal logic with the PS side. Indeed, the \staller
is the first module encountered by any CPU-originated memory request
targeting the reorganized data.
Upon arrival of a CPU-originated read transaction, the \staller extracts the
target data address and request ID (i.e., the tuple \{\addr, \id\}) before
forwarding them to the \bypass. This transfer is done through a dedicated and
unidirectional channel.
It is also the responsibility of the \staller to formulate AXI-compatible data responses based on the requested reorganized cache-line and the corresponding ID (i.e., \{\id, \data\}) as they are made available by the \bypass.
Because the CPUs can issue multiple asynchronous requests, the \staller-\bypass interface has been designed to handle multiple outstanding transactions. 


\begin{table}[t!]
  \resizebox{\columnwidth}{!}{\begin{tabular}{lcll}
   \toprule
    \textbf{Parameter} & \textbf{symbol} & \textbf{Address} & \textbf{Description} \\
  	\toprule
    Row size              & $R$          & \texttt{base+0x00}         & database tuple width\\
    Row count             & $N$          & \texttt{base+0x04}         & database tuple count\\
    Software reset        & $SW$         & \texttt{base+0x08}         & software triggered reset request\\
    Enabled columns count & $Q$          & \texttt{base+0x0c}         & amount of columns of interest\\
    Column width          & $C_{A_j}$    & \texttt{base+0x10+(j*0x2)} & $j$-th column width ($j \in [0, 11[$)\\
    Column offset         & $O_{A_j}$    & \texttt{base+0x26+(j*0x2)} & $j$-th column offset ($j \in [0, 11[$)\\
    Frame number          & $F$          & \texttt{base+0x3c}         & filtered table frame number\\
  	\bottomrule
  \end{tabular}
  }
  \caption{\sysName configuration port: addresses and description.}
  \label{tab:configuration-port}
  \vspace*{-0.39in}
\end{table}

\Paragraph{\requestor}
The \requestor leverages the data geometry passed via the configuration interface (see Table~\ref{tab:configuration-port}) to orchestrate access to main memory. It generates a deep sequence of \emph{request descriptors} indicating, for each row, the beginning (and length) of useful data within a set of bus-width-aligned transactions. Each descriptor also indicates the positions where to store the extracted columns. The descriptor is generated such that the resulting main memory requests are always bus-width aligned and with variable burst length, to never fetch more data than strictly needed.

Internally, the \requestor keeps track of the absolute position $P_{i, j}$ at which the useful data starts for the $j$-th column of the $i$-th row, with $i \in [0, N[$ and $j \in [0, Q[$, by computing
\vspace{-0.05in}
\begin{equation}
  P_{i, j} = R \cdot i + \sum\nolimits_{k=0}^{j} O_{A_k}.
  \label{eq:req_pos}
  \vspace{-0.05in}
\end{equation}

The $i$-th descriptor generated by the \requestor is then comprised of
five parameters that depend on the constant platforms-specific bus
width $B_W$. These are (1) the main memory address $R^{addr}_{i, j}$ to
fetch the $(i,j)$-th chunk of useful data; (2) the burst length
$R^{burst}_{i, j}$ of each main memory request; (3) the position
$W^{addr}_{i, j}$ in the internal \sysName's buffer where to store the
extracted chunk of data; (4) the leading $E^{s}_{i, j}$ and (5) trailing
$E^{e}_{i, j}$ number of bytes to be discarded in the response received from
main memory. These parameters are generated as shown in
Eq.~\eqref{eq:req_raddr} through~\eqref{eq:ext_end}, where the
``$//$'' operator represents the integer division and the ``$\%$''
operator represents the remainder of an integer division.

{%
\small
\vspace{-0.15in}
\begin{align}
  R^{addr}_{i, j}  &= (P_{i, j} // B_w) \cdot B_w  \label{eq:req_raddr} \\
  R^{burst}_{i, j} &= \lceil ((P_{i, j} \% B_w) + C_{A_j}) / B_w \rceil \\
  W^{addr}_{i, j}  &= (i-1) \cdot \sum\nolimits_{k=0}^{Q} C_{A_k} + \sum\nolimits_{k=0}^{j-1} C_{A_k}\\ 
  E^s_{i, j}       &= P_{i, j} \% B_w \\
  E^e_{i, j}       &= (P_{i, j} + C_{A_j}) \% B_w \label{eq:ext_end}
\end{align}
\vspace{-0.2in}
}

Every time a descriptor is produced, it is passed to any idle
\fetch. The data obtained by the \fetchs when processing each request
descriptor is stored in the \relcache. The \requestor can interface
with multiple \fetchs to decouple the request descriptor generation,
the actual memory fetches, and the extraction of relevant data. This
way, it is possible to quickly scale up the number of available
\fetchs to achieve higher memory parallelism.
In case all the \fetchs are busy, the \requestor stalls and waits for any
\fetch to become available.
Ideally, the amount of \fetchs should maximize the utilization while keeping the target memory controller below its saturation point.

When requesting data from main memory, only those locations containing
valuable data at the granularity of the bus width are accessed. In
order to access the desired data efficiently, the \requestor produces
descriptors that instruct a given \fetch to perform variable-length
memory bursts---via the $R^{burst}_{i, j}$ parameter. This contrasts with a
generic cache controller that always uses the same burst length
required to fetch an entire cache line.



        



\Paragraph{\fetch}
Each \fetch is responsible for retrieving one fixed-size chunk of data
from main memory and then directing it to the designated place in
the \relcache. The unit is internally structured in several
distinguished sub-component, namely the \reader, the \extcol, and
\writer, briefly described below.

The \reader directly interacts with the main memory controller. It
primarily produces memory fetch requests that reflect the
specifications of the descriptor passed by the \requestor.
The \reader uses the AXI protocol to perform variable-burst memory
requests towards main memory at the granularity of a single bus beat
(i.e., bus-width, which is typically a fraction of the cache-line size).
The data payload obtained by the \reader is passed to the \extcol
module.

The \extcol module extracts the individual bytes that correspond to
the portion of the columns of interest (Listing~\ref{lst:query-c},
lines~\ref{line:grp_str_start}--\ref{line:grp_str_end}). If the target
column(s) spans multiple bus lines, it waits until all the required
data items are accumulated and indicates the output's validity by
setting the enable signal. If necessary, it performs appropriate
shifting to pack useful data into contiguous chunks. It
finally passes the packed data to the \writer module.

The \writer receives the data from the \extcol, along with the
location where to store the packed data in the \relcache. It then
forms a single write request with said data payload towards the
\relcache. Any write operation emitted by a \writer module passes
through the \bypass.




\Paragraph{\relcache}
Two internal memory blocks serve as Scratch Pad Memories (SPMs) to
store the data and the metadata. The \textbf{Data SPM} serves as
the buffer to store individual extracted chunks of data arriving from the
\fetch and the \textbf{Metadata SPM} stores the bookkeeping information
maintained by the \bypass. For each cache-line, the latter stores the tuple \{$P$, $K$, \id\}, where
(1) $P$ is the epoch to which the line belongs to;
(2) $K$ tracks the number of valid bytes in the
cache-line; and (3) \id~represents the stalled transaction ID. The \id~is non-null only if
the data is incomplete and a transaction requesting the line is pending.

As the data lines are being filled with data coming from the \fetchs, the
valid byte count $K$ is incremented. The associated data line is considered 
\emph{complete} once the count reaches the cache-line size (i.e., 64~bytes).
When this happens, $K$ is set back to 0 and $P$ is updated with the 
current \sysName epoch. A status line is interpreted the by the \bypass as 
complete iff $P$ matches the \sysName's current epoch.
In other words, a status line featuring an epoch different than the 
\sysName's current epoch is considered as incomplete.

The epoch mechanism enables quick invalidation of the content of both SPMs
because changing the \sysName's epoch makes every 
SPM entry incomplete. The software-triggered reset mechanism relies on this 
mechanism to invalidate the SPMs in a single clock cycle.

\subsection{Data-path and Work-flow}
\label{subsec:flow}
\vspace{-0.02in}

In this section, we go over the workflow of the \relacc following a
given CPU-originated read transaction. We will consider two
scenarios. First, (1) we consider the case where the requested data
has already been fetched from main memory and re-organized. Thus, it
can be immediately sent to the requesting CPU (\relcache
hit). Next, (2) we cover the case in which the target data needs to be
fetched from main memory (\relcache miss). We consider the two cases
separately and refer to Figure~\ref{fig:design} as we discuss each step
of the flow.

The accelerator is designed to be generic and to operate with various
data layouts. Therefore, the software must initially configure the
\relacc with the geometry of the target relation as described in
Section~\ref{sec:design} and depicted in Figure~\ref{fig:design}
\circled{0}.
Once a core emits a AXI read request, it is intercepted by the
\relacc. Next, \circled{1} the \staller extracts the \{\addr, \id\} fields
and \circled{2} pass them to the \bypass. The latter checks whether
the new request can be immediately
served (\relcache hit) or if it must be stalled (\relcache miss). The
check is done by using the \addr~ field to \circled{3} fetch the cache-line status
from the \textbf{Metadata SPM}. Speculatively, \circled{3} the (possibly valid) content
of the requested cache-line is fetched from the \textbf{Data SPM}.

\Paragraph{\relcache Hit}
If the cache-line was marked as complete, the \bypass can immediately send
its content---i.e., the tuple \{\id, \data\}---to the \staller \circled{4}.
Then, using this information, the \staller forms an
AXI compliant transaction to reply to the CPU's initial request \circled{5}.

\Paragraph{\relcache{} Miss}
If part of the data composing the requested cache-line is missing,
the request must be stalled. In this case, the request \id~ is stored in the metadata SPM.
Once enough data returns from the \fetchs and the cache-line becomes complete, the \id~ is 
removed and the \{\id, \data\} tuple is sent to the \staller \circled{4}. If the miss in 
question is the first miss of the frame, a signal is sent the \requestor 
module to start the descriptor generation.

As described in Section~\ref{sec:design}, the \requestor has a crucial role
in orchestrating the \fetchs and
their interaction with main-memory, data extraction, and data
forwarding to the \relcache. The \requestor prepares a series of
descriptors for the \fetchs using Eq.~\ref{eq:req_pos}--\ref{eq:ext_end}.
It keeps track of which \fetch is currently busy and provides the next 
descriptor to any available \fetch \circled{B}.

Upon receipt of a new descriptor, the \fetch sends a request for a
burst of $R^{burst}_{i, j}$ data responses towards main memory at location
$R^{addr}_{i, j}$ \circled{C}. Once the full response is received, the
\extcol performs data filtering using the parameters $E^{s}_{i, j}$ and
$E^{e}_{i, j}$. Next, the filtered data is sent to the \relcache at the
address specified by $W^{addr}_{i, j}$. On its 
way to the \relcache, the filtered data chunks go through the \bypass
\circled{D}. The latter simultaneously updates the record of the newly
(partially) filled cache-lines in the Metadata SPM \circled{E}. More
importantly, the \bypass recognizes if the recent update was the last
missing part of any incomplete cache-line. In other words, by
simultaneously fetching both metadata and data record of the most
recently updated cache-line \circled{3}, the \bypass checks for any
full cache-line. In which case, the \bypass immediately sends the
corresponding \data back to the \staller \circled{4} to de-queue any
pending \{\addr, \id\} request. The \staller then replies to the CPU
by forming a \{\id, \data\} response \circled{5}.


\subsection{Toward Efficiency and Parallelism}
      \label{subsec:versions}
      The memory subsystem is a
      well-known performance bottleneck in modern SoCs.
      This is due to the characteristic latency of
      the main memory cells (e.g. DRAM) or due to the constrained bandwidth between PL and main memory.
      Common strategies for taking full advantage of memory subsystems include (1)
      making memory accesses more efficient by re-ordering the
      requests to benefit from an already-open DRAM row, and (2)
      improving DRAM bank-level parallelism to improve
      throughput.  Taking inspiration from the mentioned techniques, we consider
      the design described in Section~\ref{subsec:flow} as
      the baseline. We have then implemented a number of revisions
      to our design with increasing level of refinement to improve memory parallelism.

      We consider the design presented earlier
      in this section as our \emph{Baseline Design}
      (or \textbf{BSL}).
      The first revision of the BSL design consists in the
      integration of an intermediate buffer inside the \fetch. The rationale behind
      this modification is that a sizable amount of time is lost
      when the data is being written to the BRAM once it has been read
      and filtered by the \reader and the \extcol. By introducing a simple
      register in charge of \emph{packing} the filtered columns and only writing their
      content to the BRAM once a full cache-line has been extracted, we reduce the amount of
      accesses to the \relcache. Following the addition of this \emph{packer}, we identify this revision as \textbf{PCK}.
      The second and final revision of the design aims at
      increasing the utilization of the bus between the \reader and the
      DRAM controller by enhancing the \emph{Memory Level Parallelism} capability
      of the design. Hitherto only one outstanding read transaction was supported by
      the \reader whereas, in the \textbf{MLP} revision, the \fetch
      has been augmented to emit up to 16 independent outstanding transactions.


\newcommand{\rev}[1]{#1}

\section{Evaluation}\label{sec:eval}
\vspace{-0.02in}
We now show experimentally that \sysName offers efficient native accesses 
of any group of columns, outperforming direct row-wise accesses, and in
most cases also columnar accesses.

\subsection{Target Platform}
\vspace{-0.02in}
We implement \sysName using a Xilinx Zynq UltraScale+ MPSoC platform~
\cite{zynq_ultrascale}. The development board, Xilinx 
UltraScale+ ZCU102, is equipped with 4 Cortex-A53 cores each associated with a 
private 32~KB L1 data cache and grouped together by a shared unified L2 cache 
of 1~MB. On the PL-side, \sysName employs two internal memory blocks: the 
\textbf{Data SPM} and the \textbf{Metadata SPM}. The former, a large chunk of 
2~MB directly impacts the critical path and prevents the design from reaching 
higher frequencies. For that reason, the presented design is constrained to 
100~MHz (i.e., one-third of the maximum reachable frequency). 
The PL-side 
resource utilization is discussed in Section~\ref{subsec:pl-reports}. 
The benchmark queries are compiled with GCC 7.3.1 for 
ARM64 and executed on Linux 4.14.

\subsection{Relational Memory Benchmark}
\vspace{-0.02in}
\label{sec:eval-benchmark}
We designed a synthetic benchmark to test the behavior of \sysName under a number of representative query access patterns. The benchmark,
termed \emph{Relational Memory Benchmark}, consists of six queries focusing on projection, selection, and
aggregation. The queries are executed assuming that all data is in main memory. The queries are shown in Listing~\ref{lst:query1}.
$Q0$ is the simplest query that calculates an aggregate of a single column.
$Q1$ is a projection of $k$ columns (non-contiguous or contiguous), where $k$ can be varied.
$Q2$ projects one column and imposes a selection condition on a second column.
$Q3$ performs an aggregation (sum) over a subset of column, selected based on a predicate on another column.
$Q4$ further generalizes $Q4$ by adding a group by statement based on a third column.
Finally, $Q5$ performs a hash join query over two distinct tables.

In all the queries except for $Q5$, the benchmark only accesses a relation $S$ with $n$ columns $A_1,\dots,A_n$. Each column $A_i$ has a tunable width $C_{A_i}$. $Q5$ also accesses a second table $R$ with similar structure. 

\begin{lstlisting}[language=SQL, numbers=none, caption={Queries 1-6}, label={lst:query1}]
  Q0: SELECT SUM(A1) FROM S;
  Q1: SELECT A1, A2, ..., Ak FROM S;
  Q2: SELECT A1 FROM S WHERE A3 > k;
  Q3: SELECT SUM(A2) FROM S WHERE A4 < k;
  Q4: SELECT AVG(A1) FROM S WHERE A3 < k GROUP BY A2;
  Q5: SELECT S.A1, R.A3 FROM S JOIN R ON S.A2 = R.A2;
\end{lstlisting}

\Paragraph{Benchmark \& Measurement Setup} Unless otherwise stated, the row-size of the
benchmark data is 64 bytes and the column-size is 4 bytes. Throughout
our experimentation we test with varying values of both to see their 
impact as well. Further, the data size is by default 32MB and we increase 
it up to 2GB for the scalability experimentation. When reporting latency
numbers, we avoid measurement anomalies by repeating each experiment $30$ 
times and reporting averages and standard deviations.

\subsection{Experimental Results}
\vspace{-0.02in}

\begin{figure}[t!]
  \centering
  \includegraphics[width=1.0\linewidth]{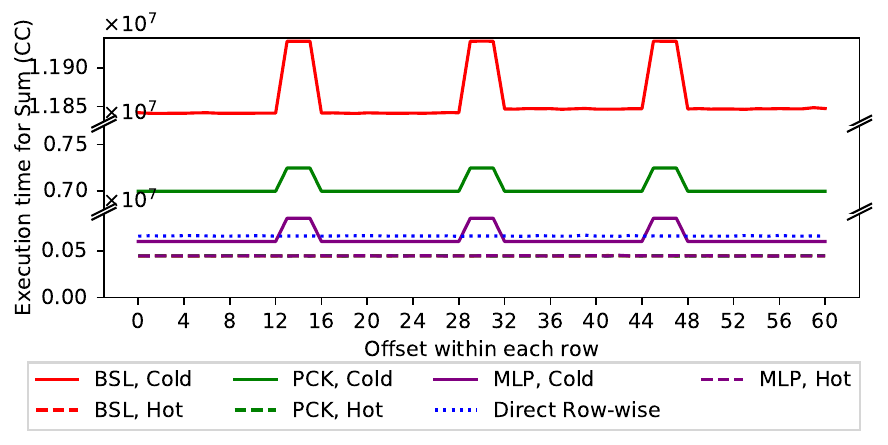} 
  \vspace{-0.28in}
  \caption{The fastest hardware revision (\emph{MLP}), \sysName outperforms direct accesses to row-oriented memory. The projected column's offset does not impact \sysName's performance.}
  \label{fig:offset_in_queries}
  \vspace{-0.1in}
\end{figure}

\Paragraph{Impact of Hardware Revisions}
Our first experiment shows the impact of the different hardware revisions
introduced in Section~\ref{sec:design}. 
Figure~\ref{fig:offset_in_queries} shows the execution time of $Q0$ where 
we calculate the sum over one column. The $x$-axis varies
the offset of the projected column, and the seven different lines 
correspond to the following configurations: three \sysName versions 
(\textbf{BSL}, \textbf{PCK}, \textbf{MLP}, see 
Section~\ref{subsec:versions}) either \emph{cold} or \emph{hot} and 
direct DRAM access. 
We observe that there is a progressive performance improvement from our hardware revisions. 
Initially, cold \textbf{BSL} is 16$\times$ slower than loading data directly from a
row-oriented layout from main memory due to the lack of memory parallelism.
On the contrary, with our most optimized revision (\textbf{MLP}) even 
without having the projected column in the Reorganization Buffer (\textbf{MLP} \emph{cold}, shown in the solid purple line), \sysName
is faster than a direct access in the row-oriented data in memory (blue dotted line). 
Note that all hot (dashed) lines coincide irrespectively of the hardware
revision.
Further, we observe that projecting \emph{cold} columns 
(i.e., the requested columns are not initially buffered in the \relcache)
incurs up to $\boldsymbol{20\%}$ 
lower latency than going through the normal route for our \textbf{MLP} design 
(purple line in the figure). The reason is that (1) \sysName better exploits the internal 
memory bandwidth to fetch only the desired data items at bus-width granularity, and (2) the CPU caches are not polluted with
unwanted fields and are therefore used more efficiently.

In the rest of our experimentation, we use the \textbf{MLP} 
hardware revision. As a result, in the remainder of this section, when
using \sysName we refer to the performance of the \textbf{MLP} hardware
revision.

\Paragraph{Column Offset does not Impact Performance} 
Figure~\ref{fig:offset_in_queries} has a second important message. By
varying the offset $O_{A_1}$ of the projected column $A_1$, we identify
that the offset from which we are reading from, generally does not affect 
performance. The experiment considers a table of 64-byte rows and a 
target 
column width $C_{A_1}$ = 4~bytes. We observe that the value of the offset 
has no impact on latency, especially when DRAM is directly accessed and 
when \sysName is \emph{hot}, regardless of the considered revision. 
We observe three spikes (at 13 to 15, 29 to 31, and 45 to 47) for the \emph{cold} cases (i.e., when the targeted data is not yet ready in the \relcache).
The reason behind these spikes is that as we increase $O_{A_1}$, the descriptors emitted by the \requestor vary.
In fact, most of the time, the 4 bytes of interest fit within 16~bytes (i.e., the width of the bus), leading the \requestor to create a read transaction with a burst length of 1.
However, when the offset plus the size of the data does not fit in a single bus width (e.g., 13+4, 29+4, 45+4, \ldots), the \requestor must emit read requests with a burst length of 2, leading to a slightly higher latency.

Since the offset of the targeted column does not affect 
query performance, in the remainder of our experiments, the first
targeted column will always be $A_1$ that starts at the beginning of the 
row.

\vspace{-0.00in}
\begin{figure}[t]
  \centering
  \includegraphics[width=1.0\linewidth]{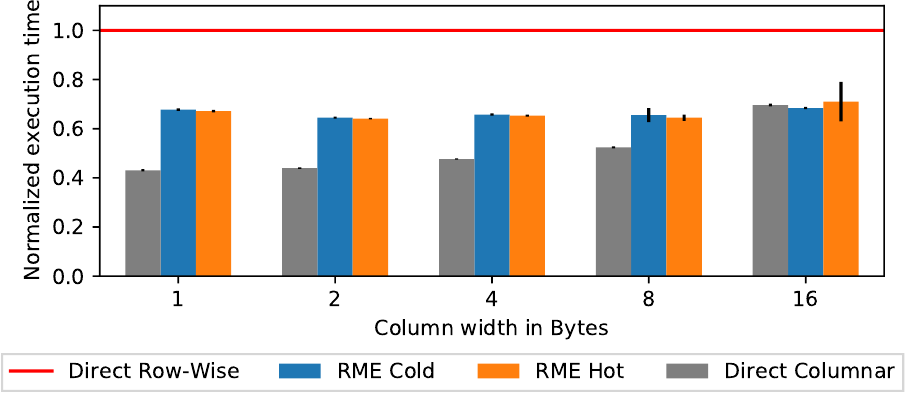}
  \vspace{-0.26in}
  \caption{The normalized execution time for Q1. \emph{MLP} outperforms
  row-oriented direct memory accesses, making our prototype an accelerator
  that can offer the optimal data layout for a query at a lower latency than DRAM.}
  \label{fig:Q1}
  \vspace{-0.20in}
\end{figure}


\Paragraph{\sysName Enables Native Columnar Accesses}
Our first experiment shows that \sysName can efficiently access---and
propagate through the cache hierarchy---individual columns when
reading row-oriented data. We now evaluate $Q1$ with $k = 3$.
The three target columns are not contiguous, with offsets $O_{A_1} = 0$, $O_{A_2} = 24$, and $O_{A_3} = 24$ (i.e. $A_3$ has offset 0+24+24=48 from the beginning of the row) respectively. All three columns 
have same width which varies in the range [1,16]~bytes.
Figure~\ref{fig:Q1} shows the normalized execution time to complete $Q1$.
We compare the time to access the data
directly from the row-oriented organization in main memory
(\emph{Direct Row-wise} access) and through our \sysName. 
We consider \textbf{MLP} with both \emph{hot} and \emph{cold}
accesses. 
Finally, we also compare against direct access to data already structured in columnar format (\emph{Direct Columnar} access).


Figure~\ref{fig:Q1} shows that \sysName outperforms direct row-wise access irrespectively of whether accesses are cold or hot. 
The takeaway is twofold. First, \emph{accessing a 
group of columns via \sysName} delivers the data with the \emph{optimal
layout and outperforms direct accesses} to a row-oriented version of 
data in memory. Second, \emph{\sysName achieves an average latency
that is comparable to pure columnar accesses}. Specifically, 
Figure~\ref{fig:Q1} shows that for column size 16~bytes and above, 
$Q1$ is faster through \sysName rather
than through a pure column-store. Therefore, data can be simply stored
row-wise in memory while any hybrid layout can be delivered by the \sysName 
with no row-to-column data transformation latency.



\begin{figure}[t]
  \centering
  \includegraphics[width=1.0\linewidth]{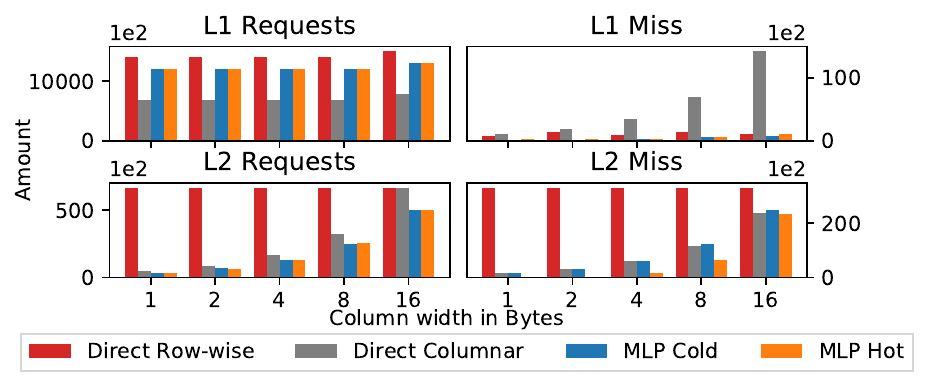}
  \vspace{-0.25in}
  \caption{L1 and L2 cache requests and misses during the execution of Q1.
  The \sysName can achieve higher hit ratio since it re-organizes the target column in
  an optimal layout.}
  \label{fig:cache-miss}
  \vspace{-0.22in}
\end{figure}

\Paragraph{Cache Miss Ratio}
The benefits from \textbf{MLP} observed in Figure~\ref{fig:Q1} for $Q1$ can be further
explained if we take a careful look at the composition of cache requests and misses.
Figure~\ref{fig:cache-miss} shows the L1 (top) and L2 (bottom) cache
requests/misses for the \textbf{MLP} design. We observe that \sysName causes better cache utilization in both
L1 and L2, which explains the significant performance savings despite
the cost of routing data through the accelerator. Note that the total size 
of the target column does not fit in L1, but since the column is accessed sequentially, 
the L1 pre-fetcher can drastically reduce the L1 misses. The high number of L2 requests 
is attributed to the L1 pre-fetcher. Overall, \sysName provides ideal cache locality as no
extra data items are ever propagated to the CPU caches beyond what strictly
required to execute the query at hand. 
Conversely, accessing the three columns in the column-store creates three
independent sequential access streams. The CPU in our target SoC supports up to
four independent pre-fetch streams. Therefore most of the memory access cost is successfully hidden.
On the other hand, directly accessing a few columns from a row-wise data 
representation in memory fetches a lot of unnecessary (columns), leading 
to poor locality.

\begin{figure}[t!]
  \centering
  \includegraphics[width=0.97\linewidth]{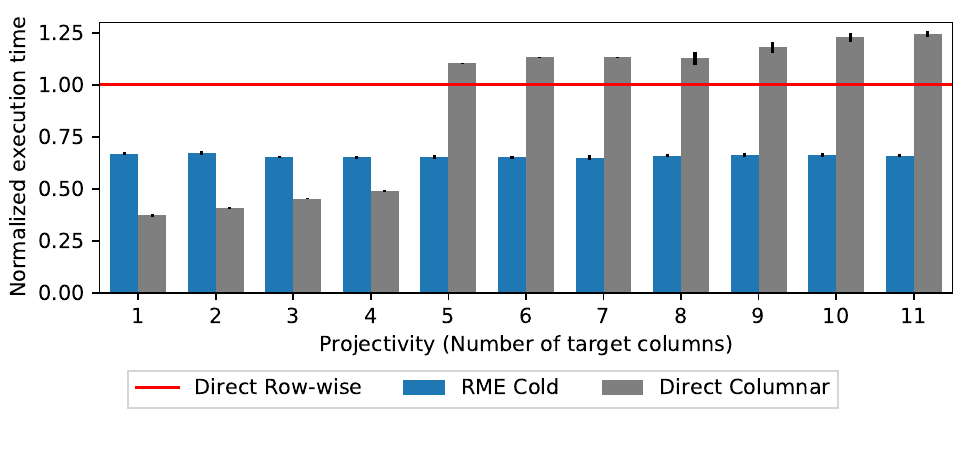}
  \vspace{-0.25in}
  \caption{\sysName has stable performance irrespectively of projectivity when compared with row-wise memory accesses. Further, \sysName outperforms columnar accesses when projecting more than 4 columns in $Q1$.}
  \label{fig:projectivity}
  \vspace{-0.25in}
\end{figure}

\Paragraph{\sysName Has Stable Performance as Projectivity Increases}
When comparing with columnar accesses we have to also take into account
the tuple materialization cost. In our next experiment we vary the
projectivity from 1 to 11 columns. Figure~\ref{fig:projectivity} shows
that for low projectivity (between 1 and 4) reading from a columnar
database is faster than \sysName. For projectivity of more than 5 
columns, \sysName outperforms direct columnar accesses because of the 
tuple reconstruction cost. In addition, through our profiling we
observed that the prefetecher can recognize up to four parallel 
sequential streams of accesses, which helped the columnar accesses for
low projectivity. Overall, \sysName consistently outperforms direct
row-wise accesses that pollute the caches with unwanted fields and 
outperform columnar access beyond a projectivity threshold.


\Paragraph{Setup for $Q2$ through $Q5$}
We continue our experimentation with $Q2$ through $Q5$ from
our benchmark from Section~\ref{sec:eval-benchmark}, focusing on the comparison between RME and direct row-wise access.
Two sets of experiments are conducted for each query. First, we vary
the column size of a table with 64~byte-wide rows; second, we access 
4~byte-wide columns while varying the row size. The default cardinality
of the base table is 44K rows.


\begin{figure*}
  \centering
  \begin{subfigure}[b]{0.33\textwidth}
    \centering
    \includegraphics[width=\linewidth]{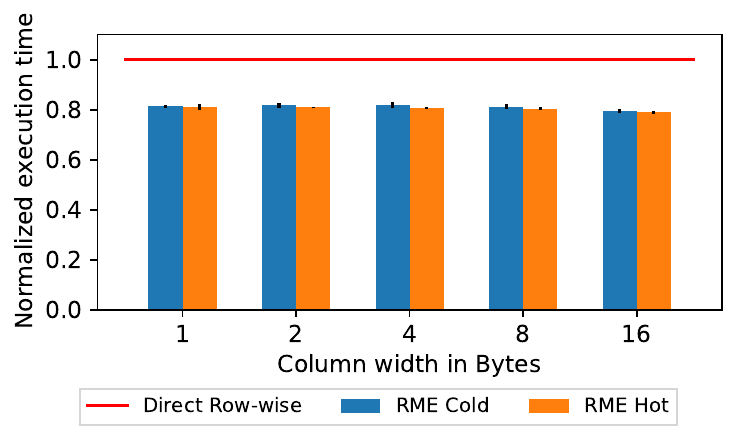}
    \vspace{-0.22in}
    \caption{Query 2}
    \label{fig:q2-col}
  \end{subfigure}
  \hfill
  \begin{subfigure}[b]{0.33\textwidth}
    \centering
    \includegraphics[width=\linewidth]{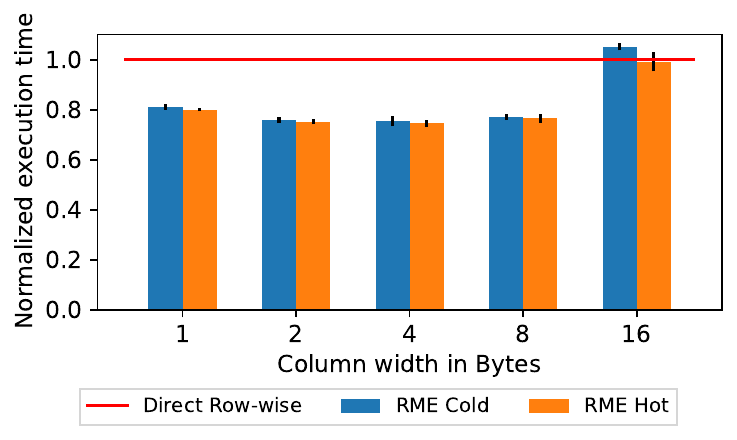}
     \vspace{-0.22in}
    \caption{Query 3}
    \label{fig:q3-col}
  \end{subfigure}
  \hfill
  \begin{subfigure}[b]{0.33\textwidth}
    \centering
    \includegraphics[width=\linewidth]{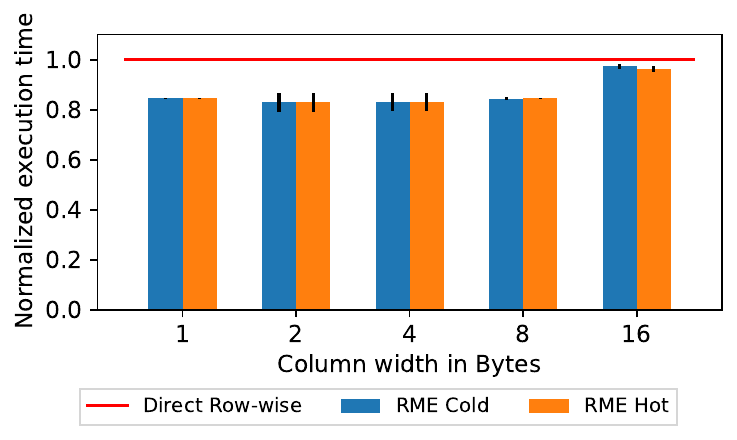}
        \vspace{-0.22in}
    \caption{Query 4}
    \label{fig:q4-col}
  \end{subfigure}
  \vspace{-0.28in}
  \caption{Aggregation queries with varying column size.
  Depending on the query, the benefit of using
  \sysName varies. However, \sysName outperforms row-oriented direct memory
  accesses since it accesses only useful data.
  }
  \label{fig:aggregation-col}
  \vspace{-0.05in}
\end{figure*}

\begin{figure*}
  \centering
  \begin{subfigure}[b]{0.33\textwidth}
    \centering
    \includegraphics[width=\linewidth]{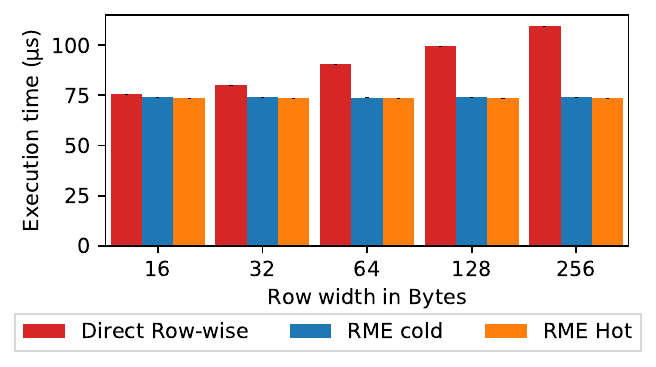}
    \vspace{-0.22in}
    \caption{Query 2}
    \label{fig:q2-row}
  \end{subfigure}
  \hfill
  \begin{subfigure}[b]{0.33\textwidth}
    \centering
    \includegraphics[width=\linewidth]{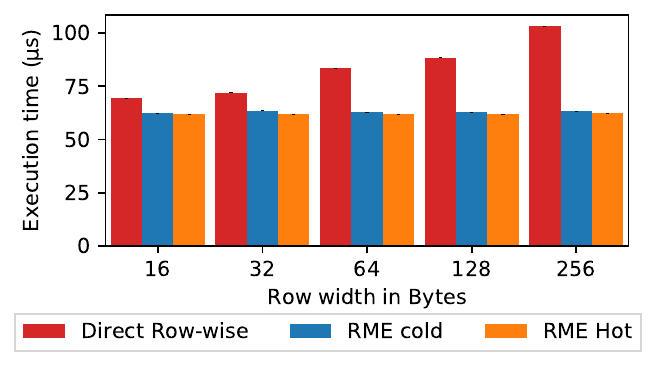}
        \vspace{-0.22in}
    \caption{Query 3}
    \label{fig:q3-row}
  \end{subfigure}
  \hfill
  \begin{subfigure}[b]{0.33\textwidth}
    \centering
    \includegraphics[width=\linewidth]{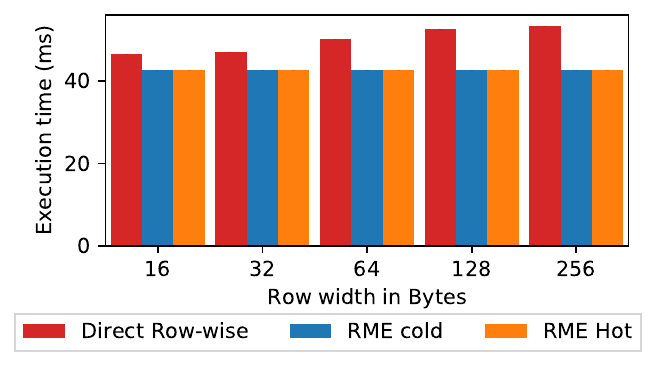}
        \vspace{-0.22in}
    \caption{Query 4}
    \label{fig:q4-row}
  \end{subfigure}
    \vspace{-0.28in}
  \caption{Aggregation queries with varying row size.
  Depending on the query characteristics and projectivity, the
  benefit varies. However, \sysName enables more efficient accesses by providing
  the optimal layout.}
  \label{fig:aggregation-row}
  \vspace{-0.12in}
\end{figure*}

\Paragraph{\sysName Offers Efficient Near-Memory Projection}
We now discuss the performance of $Q2$ that has around 90\% selectivity.
$Q2$ benefits by fetching only the two desired columns instead
of the entire row, while the selection of $Q2$ takes place on the software side. The performance graph is shown in Figure~\ref{fig:q2-col}. We observe that \sysName offers faster execution
time in both \emph{cold} and \emph{hot} cases. 
Figure~\ref{fig:q2-row} shows that the performance gain of \sysName
increases for larger row size (up to 1.4$\times$). 
We note that \emph{\sysName's latency remains virtually the same} as it accesses
exactly the desired amount of data. However, answering the query via the
direct access of the row-oriented data leads to poor cache utilization as
larger rows lead to higher cache pollution.
This shows that \sysName has a much stable and predictable performance irrespective of the row size.

\Paragraph{Selection, Projection, and Aggregation Queries}
In our next experiment, we consider the more complex queries $Q3$ and
$Q4$, that test selection, projection, aggregation, and group by. The 
selectivity of $Q3$ and $Q4$ is less than 10\%.
Similarly to before, we stress the \sysName using
two different sets of experiments.
Figures~\ref{fig:q3-col} and~\ref{fig:q3-row}
show the normalized latency of $Q3$ when varying the column
size with fixed row size, and when varying the row size with fixed column 
size, respectively. When the \sysName
is used, the execution time is faster than the traditional data access to main memory.
For $Q4$, in particular, the cost of group by dominates the execution time compared to 
data accessing, thus, the performance improvement is reduced as shown in Figures~\ref{fig:q4-col} and~\ref{fig:q4-row}. 
We note that both $Q3$ and $Q4$ have a performance drop for column width
16~bytes. This is attributed to the fact that in some cases we need to 
fetch data spanning 
32-bytes, i.e., half the cache line size. This happens more frequently 
due to the low selectivity. As a result, the 2$\times$ 
increase in efficiency in the cache utilization is canceled out by the 
overheads of routing through PL memory.



\begin{figure}
  \centering
  \begin{subfigure}{.47\linewidth}
    \centering
    \includegraphics[width=\linewidth]{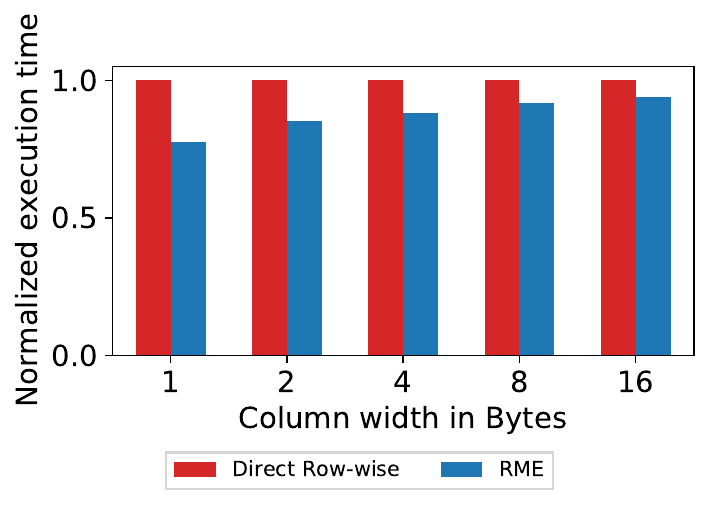}
    \vspace{-0.21in}
    \caption{Varying column size}
    \label{fig:join}
  \end{subfigure}%
  \begin{subfigure}{.47\linewidth}
    \centering
    \includegraphics[width=\linewidth]{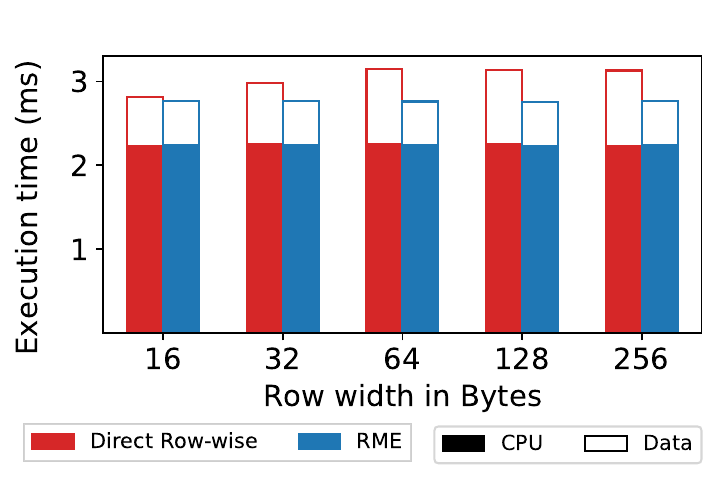}
    \vspace{-0.21in}
    \caption{Varying row size}
    \label{fig:join_r}
  \end{subfigure}
  \vspace{-0.1in}
  \caption{\sysName performs join faster than traditional row-store join by minimizing data movement.}
    \vspace{-0.25in}
\end{figure}

\Paragraph{\sysName Reduces Data Movement for Joins ($Q5$)}
When considering queries that join multiple tables ($Q5$), \sysName 
can help to project only the relevant columns, that is, the columns of
the join attributes and the column(s) projected in the SELECT statement
of the query. In this experiment, we join using a state-of-the-art
hash-based join algorithm with a single-pass hash table generation, which
is then probed by the second relation. Half of the entries of the outer
relation have a match in the inner relation. Figure~\ref{fig:join} shows 
the normalized query latency while varying target column sizes. We
observe that joining through \sysName gives a benefit between 5\% and
10\%. Figure~\ref{fig:join_r} compares the execution time of this query for different row sizes. \sysName reduces the total runtime by up to 12\% depending the row width. 
The graph also shows that the CPU overhead (solid portion of the bars) of hashing constitutes the majority of the runtime which is constant across \sysName and direct row-wise access, while, \sysName can optimize the data movement by up to 41\% as the row size increases because of its lower cache misses, better strided accesses, and higher cache utilization.

\begin{figure}
  \centering
  \includegraphics[width=0.93\linewidth]{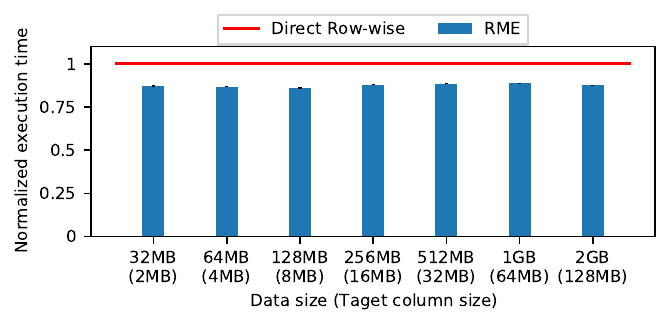}
  \label{fig:q8-col}
  \vspace{-0.1in}
  \caption{\sysName scales with data size.}
  \label{fig:frame}
    \vspace{-0.25in}
\end{figure}

\Paragraph{\rev{\sysName Scales with Data Size}}
\sysName supports arbitrary data sizes despite having a small data SPM due to the space limitations imposed by the platform.
To evaluate \sysName's scalability, we
repeat $Q1$ while projecting four target columns in larger tables ranging from 32~MB to 2~GB. Every time we fill the data SPM, we use the light-weight reset mechanism introduced in Section~\ref{sec:design}.
Figure~\ref{fig:frame} compares $Q1$ execution time with \sysName normalized by the direct row-wise access. We observe that the benefit from using \sysName remains virtually unchanged for any data size, because \sysName always provides the optimal layout.

\Paragraph{\sysName Provides Better Performance in All Queries}
Overall, the experimental analysis shows that our \sysName outperforms direct
DRAM accesses over row-oriented data, even though data accesses through
the FPGA can be quite inefficient. This result demonstrates that it is possible to
achieve native in-memory columnar accesses over data that is stored in row-oriented format.
Additional experiments that were omitted due to space constraints are available in an extended technical report \cite{roozkhosh2021relational}.


\Paragraph{Long-Term Potential and Impact}
Even though the underlying PLIM approach \cite{Roozkhosh2020} provides 
fine-grained observability and management of memory traffic between processors and 
memory hierarchy, it forces transactions to cross through a lower-frequency 
domain, i.e., that of the PL (100~MHz in our case). Because of this, an 
additional latency is added to each transaction due to the clock domain crossing 
(CDC) overhead~\cite{Hoornaert2021b,Hoornaert2021,Roozkhosh2020}. This means 
that under PLIM, the latency of individual memory transactions can be 
significantly worse than what observed with direct memory accesses. This effect 
is particularly strong in the considered hardware platform, as documented 
in~\cite{Hoornaert2021}. Nonetheless, as we observe in our experiments, the 
benefits by \sysName fully offset the described effect. This observation sets 
the basis for our long-term vision that is twofold. On the one hand, our 
\sysName is expected to provide even larger performance benefits in newer 
platforms with better PS-PL integration and lower-latency communication 
interfaces. On the other hand, the ability to offer significant performance 
advantages even at low synthesis frequencies makes our design suitable for 
integration directly within the main memory controller.

\vspace{-0.01in}
\subsection{PL Resource Utilization and Timing}
\label{subsec:pl-reports}

After the synthesis and the implementation of the design on the ZCU102
development board using Vivado 2017.4, we obtained reports
regarding the PL resources utilization of the \textbf{MLP} \sysName design.

As mentioned earlier, all the designs presented and tested in this article
run at 100~MHz. Despite being only one third of the 
achievable frequency on the target board, the \textbf{MLP} design has proved the
utilization of the resources at hand to be efficient.
As shown in Table~\ref{tab:synthesis_reports}, the area utilization never exceed 3\% except for BRAM for which 
we purposefully maxmize the size of the SPMs to improve the performance of \sysName.
The compactness of the design paves the way for more ambitious revisions where modules could be duplicated (e.g., a
design featuring multiple Fetch-Units) and pipelines extended.
Moreover, a low utilization means that proposed architecture could fit in smaller 
PS-PL platforms such as the Zybo z7-10, making our approach a good fit for edge and cloud computing.

\begin{table}[t!]
  \small
  \begin{tabular}{ccccc}
    \multicolumn{5}{c}{\textbf{Area Report}} \\
    \toprule
    \multicolumn{1}{c}{\textbf{Resources}}        & \multicolumn{1}{c}{\textbf{LUT}} & \multicolumn{1}{c}{\textbf{FF}} & \multicolumn{1}{c}{\textbf{BRAM}} & \multicolumn{1}{c}{\textbf{DSP}} \\
    \toprule
    \multicolumn{1}{c}{Utilization (\%)} & \multicolumn{1}{c}{2.78} & \multicolumn{1}{c}{0.68} & \multicolumn{1}{c}{60.69} & \multicolumn{1}{c}{0.08} \\
    \bottomrule
  \end{tabular}
  \caption{Post-implementation area report for the \textbf{MLP} design implemented on the Xilinx ZCU102.}
  \label{tab:synthesis_reports}
  \vspace{-0.4in}
\end{table}

\section{Related Work}\label{rworks}
\vspace{-0.03in}

\Paragraph{Hybrid Layouts} 
Following the \emph{one size does not fit all} rule~\cite{Stonebraker2005a}, many HTAP systems use the row-format to ingest data and then convert it to columnar-format for analytical processing~\cite{Ozcan2017}.
Examples include SAP HANA~\cite{Farber2012}, Oracle TimesTen~\cite{Lahiri2013}, MemSQL~\cite{Shamgunov2014}.
Thus, these HTAP systems fuse the data ingestion and data analytics pipelines.
The \emph{optimal} layout is more often neither a column-store or a row-store~\cite{Alagiannis2014}.
Systems like H$_2$O~\cite{Alagiannis2014}, Hyper~\cite{Kemper2011}, Peloton~\cite{Arulraj2016}, and OctopusDB~\cite{Dittrich2011} use adaptive layouts depending on the query access patterns.
For example, OctopusDB maintains several copies of a database stored in different layouts.  It thus so by means of a logical log as its primary storage structure and then creating secondary physical layouts from the log entries.
H$_2$O is another hybrid system which dynamically adapts the storage layout depending on the workload.
It materializes parts of the data in various patterns depending on the query and as the workload changes, the storage and access patterns keep adapting accordingly.
Peloton also uses an adaptive policy, however instead of an \emph{immediate} policy, it adopts an \emph{incremental} data reorganization policy.
Besides, H$_2$O uses multiple execution engines to keep the same data in different layouts, whereas Peloton uses a single execution engine.
These systems need to store multiple layouts of the data and also need to convert between formats which increases the complexity, materialization overhead and maintenance cost.

\Paragraph{FPGA in DBMS} 
FPGAs can be integrated either by using it as a filter by placing it 
between the data source and CPU or by using it as a co-processor to 
accelerate the workload~\cite{Fang2020, Istvan2020, Istvan2020a, Istvan2019}. 
In the former approach, the FPGA is used as a decompress-filter between 
the data source and the CPU to improve the effective bandwidth. 
This approach has been adopted by a number of systems like Netezza~\cite{Francisco2011}, 
Mellanox~\cite{NVDIA2020}, and Napatech~\cite{Napatech2018}. 
In contrast, in the latter approach, the FPGA can access the host 
memory directly and communicate with the CPU via shared memory, 
thus avoiding the extra copying of data from/to the device memory. 
Although this approach is still quite new, a few systems have deployed 
FPGAs as co-processors~\cite{Crockett2014, Gupta2016, Sidler2017, Salami2017}. 
Following this direction, different operators of DBMS like selection~\cite{SUN2021}, 
aggregation~\cite{Dennl2013}, compression~\cite{Qiao2018}, decompression~\cite{Fang2019}, 
sort~\cite{Zhang2016a}, groupby~\cite{Absalyamov2016}, and joins have~\cite{Xue2020,Halstead2015} been accelerated.
Another popular technique is to integrate FPGA to the CPU as an I/O 
device to accelerate database analytics especially where CPUs are the 
bottleneck. 
Contrary to the co-processor technique, the CPU and FPGA have their 
own memory in this architecture. 
The FPGA is connected to the CPU through buses. 
When the FPGA receives tasks from the CPU, it copies the data from the 
host memory to the device memory, then it fetches data from the memory, 
writes the results back to the device memory after processing, and 
finally copies the results back to the host memory. 
A number of systems like Kickfire's MySQL Analytic Appliance, dbX have 
implemented this architecture~\cite{Scofield2010, Casper2014}.

Many FPGA-based accelerators report high throughput, however, 
the low bandwidth between FPGA and host memory (or CPU) is a 
bottleneck~\cite{Kara2017}. In addition, the transfer latency of data 
from the host to FPGAs is significant. 
Thus, designing accelerators is challenging for systems with 
unpredictable memory access patterns. 

Another interesting line of work is query accelerators in the network layer~\cite{Aguilera2019, Amaro2020, Sidler2020, Korolija2021, AmazonAQUA}, that access non-local memory. However, 
there are common aspects with database systems that these works aim to 
reduce the number of network traversals to access non-local memory, the 
data movement inefficiencies, and network overhead by accelerating operators. 
Particularly, Farview~\cite{Korolija2021} accelerates a wide range of 
operators, including  projection, selection, aggregation, grouping, and 
encryption/decryption. 

In contrast to the abovementioned approaches, we present a completely 
new approach where we transparently transform data from rows to columns 
with the help of re-programmable logic.
Our approach does not require to copy additional data, has minimal 
runtime overhead, and shows promising performance.


\vspace{-0.01in}
\section{Conclusion and Future Work}\label{conc}
\vspace{-0.01in}



In this paper, we present a radically new approach to offer access to 
both row-oriented and columnar layouts. We build on recent developments in
reprogrammable hardware to implement logic between the memory and the 
processor, which is able to on-the-fly convert rows to arbitrary groups 
of columns. Our approach, named \emph{Relational Memory} pushes projection
from software to hardware and allows the same software implementation to have
native access to both row-oriented and column-oriented data layouts.

Our prototype implementation demonstrates this functionality and offers
access to arbitrary groups of columns at no additional latency than
accessing directly the optimal data layout. This groundbreaking result
opens a new avenue for software/hardware co-design of data systems since
it allows a single code path to have access to the desired data layout for
each query. 
 
Overall, \emph{Relational Memory} is the first step to a new class of data 
systems architectures, as implementing projection in hardware lays the 
groundwork for other relational operators (selection, aggregation, 
group by, join pre-processing). Finally, the low-end hardware used in
our prototype underlines that it is feasible to integrate this logic in 
memory controllers widening its impact.



\balance

\bibliographystyle{ACM-Reference-Format}
\bibliography{library,main}


\end{document}
\endinput